\documentclass[aps,prl,twocolumn]{revtex4}

\usepackage{amsmath}
\usepackage{amssymb}
\usepackage{natbib}
\usepackage{graphicx}
\usepackage{bm}

\newcommand{\sgn}{\textrm{sgn}}

\renewcommand{\Re}{\textrm{Re}}
\renewcommand{\Im}{\textrm{Im}}

\def\be{\begin{equation}}
\def\ee{\end{equation}}
\def\ba{\begin{eqnarray}}
\def\ea{\end{eqnarray}}
\def\rr{{\bf r}}

\def\w{\omega}

\def\e{\epsilon}

\def\p{\partial}
\def\jj{\textbf{j}}

\def\rr{\textbf{r}}
\def\qq{\textbf{q}}

\def\EE{\textbf{E}}
\def\BB{\textbf{B}}

\def\d{\delta}

\def\s{\sigma}

\def\bra{\langle}
\def\ket{\rangle}
\def\nbl{\bm \nabla}

\begin{document}

\title{Dynamic chiral magnetic effect and Faraday rotation in macroscopically disordered helical metals}
\date{\today}
\author{J. Ma and D. A. Pesin}
\affiliation{Department of Physics and Astronomy, University of Utah, Salt Lake City, UT 84112 USA}
\begin{abstract}
We develop an effective medium theory for electromagnetic wave propagation through gapless non-uniform systems with dynamic chiral magnetic effect. The theory allows to calculate macroscopic-disorder-induced corrections to the values of optical, as well as chiral magnetic conductivities. In particular, we show that spatial fluctuations of optical conductivity induce corrections to the effective value of the chiral magnetic conductivity. Experimentally, these corrections can be observed as sharp features in the Faraday rotation angle near frequencies that correspond to the bulk plasmon resonances of a material. Such features are not expected to be present in single crystal samples.
\end{abstract}
\maketitle
%
%
%
%
%
%
%

Ignited by the field of topological insulators, the interest to geometric properties of band structures has spread to gapless systems by now. Among the latter, Weyl semimetals seem to have attracted the largest attention, partly due to their non-trivial topological properties~\cite{WanVishwanath2011,Burkov2011,Turner2013, Hosur2013,Vafek,Burkov-Review}, partly due to the experimental verification of their existence~\cite{WSM-Exp-1,WSM-Exp-2,WSM-Exp-3,WSM-Exp-4,WSM-Exp-5,WSM-Exp-6,WSM-Exp-7,WSM-Exp-8}.

There have been a substantial number of theoretical proposals on how the geometric properties of Weyl metals manifest themselves in observable experimental quantities related, for instance, to magnetotransport~\cite{Nielsen1983,Son2013a,BurkovMR,Spivak2016,Andreev2016}, non-local transport~\cite{ParameswaranPesin2014,ParameswaranBerg}, or strain~\cite{Vozmediano2015,PikulinWSstrain,Grushin2016,Vozmediano2016} phenomena. However, in Weyl systems, or gapless topological systems in general, one necessarily deals with systems with gapless bulk. This means that at least in principle, they manifest all responses pertinent to a more mundane metal with the same symmetries. This implies that careful quantitative understanding of various experiments is required in order to disentangle the geometric features of the observed responses. In particular, the omnipresent disorder effects must always be carefully studied.


In this paper, we describe how macroscopic sample inhomogeneities affect optical tests of the \textit{dynamic} chiral magnetic effect via Faraday rotation measurements. We show that in thin films of metals with low carrier concentration, macroscopic fluctuations of local conductivity affect the frequency dependence of the measured optical polarization rotation signal, creating sharp features near the plasma edge of the metal, which are absent in single crystals.

The chiral magnetic effect (CME) is defined as the existence of a contribution to the electric current density, $\jj$, driven by a magnetic field, $\BB$, which yields the following expression for the electric current density in the simplest isotropic case:
\begin{equation}\label{eq:currentgeneral}
  \jj(\w)=\sigma(\w) \EE(\w)+\gamma(\w) \BB(\w).
\end{equation}
The first term on the right hand side of Eq.~\eqref{eq:currentgeneral} represents the usual optical conductivity response to an electric field $\EE$. The $\w\to 0$ limit of the coefficient $\gamma(\w)$, which can be non-zero in a metal, is known as the chiral magnetic conductivity in the literature. Here, we consider a more general case of frequency-dependent $\gamma(\w)$, keeping the name of the chiral magnetic conductivity for it.

There are two basic types of the CME, pertaining to the cases of purely static, and slowly oscillating $\BB$-field, which are appropriately called \textit{static} and \textit{dynamic} CME, respectively.

The static CME is of purely topological origin, and relies on the existence of Weyl points, and Berry curvature monopoles associated with them, in a band structure~\cite{Vilenkin,Cheianov1998,Nielsen1983,Kharzeev2009,SOn2013,Chen2013}. However, the static CME does not occur in equilibrium crystals~\cite{Niu2013,Franz2013}: it requires an imbalance between the chemical potentials near Weyl points with opposite signs of Berry monopole charges. This imbalance is in general hard to achieve, but when it is reached via the chiral anomaly, the static CME manifests itself either as the negative longitudinal magnetoresistance~\cite{Nielsen1983,SonSpivak2013,BurkovMR}, or non-local voltages in thin film samples~\cite{ParameswaranPesin2014}. In this sense, the static CME has been observed via magnetotransport measurements in~\cite{Kim2013,Xiong2015,zhang2016signatures,Zheng2016} (see Ref.~\cite{Hassinger2016,xu2016review} for further references and review of recent results), and via non-local voltage measurements in~\cite{Xiu2015}.

Here we focus on the dynamic CME, which does exist in equilibrium gyrotropic metals, and describes their linear response to slowly oscillating electromagnetic fields. Its low-frequency limit is of geometric origin: It comes from the local geometry of electronic bands, rather than their topology, and is due to the existence of orbital magnetic moment of quasiparticles in systems with non-zero Berry curvature~\cite{MaPesin2015,Zhong2016,AlaviradSau2016}. It also does not require the existence of Berry monopoles, but tends to be large when the monopoles are present~\cite{ChangYang2015nonode,MaPesin2015}.

Physically, the dynamic CME is a particular manifestation of the natural optical activity phenomenon~\cite{MaPesin2015,Zhong2016}. This observation prompts an experimental measurement of the chiral conductivity $\gamma(\omega)$ by studying the Faraday rotation of polarization of light transmitted through a slab of gyrotropic material.

The (complex) polarization rotation angle is determined solely by the phase difference accumulated by the two circular polarizations of light as they travel through the bulk of the material~\cite{Bassiri1988}:
\begin{equation}\label{eq:theta}
  \theta(\w)=\frac{\mu_0}2 \gamma (\w) d,
\end{equation}
where $d$ is the thickness of the slab in the propagation direction. The rotation angle is not affected by possible surface conduction, either~\cite{MaPesinunpublished}. Therefore, the Faraday rotation appears to be the most direct way to measure $\gamma(\omega)$.


Here, we show that macroscopic inhomogeneities make the effective macroscopic observable $\gamma_{\textrm{eff}}(\w)$ different from its value predicted by the band structure calculations, $\gamma_{BS}(\w)$. In particular, $\gamma_{\textrm{eff}}(\w)$ has sharp features around the plasma edge of the metal, which is~\emph{not} expected for $\gamma_{BS}(\w)$. Instead, at frequencies large compared to the inverse momentum relaxation time on the Fermi surface, and small compared to the lowest interband splitting at the Fermi surface, $\gamma_{BS}(\w)$ is a real frequency-independent constant~\cite{MaPesin2015}. More generally, when the frequency of the incident light is not small compared to relevant band splittings, $\gamma_{BS}(\w)$ does depend on $\w$, but obviously is still not expected to have any features at the plasma edge of a metal.

In what follows, we set out to construct the effective medium theory for a macroscopically disordered sample with CME. The effective medium theory for composite materials, and metals in particular, has been developed over the past century~\cite{dykhne1971conductivity,Ashcroft1977,landauer1978electrical,RaikhvonOppen2005}, but it has not been constructed for metals with natural optical activity. We fill this void below.

\textit{General formalism --} We assume that a non-uniform sample is characterized by \textit{macroscopic} inhomogeneities, which occur on length scales large compared to the microscopic ones, like the Fermi wave length, or elastic mean free path. The sample is then characterized by a space-dependent (optical) conductivity, $\sigma_{ab}(\rr,\w)$, and the CME tensor, $\gamma_{abc}(\rr,\w)$.

If the variation of electromagnetic fields is slow on the scale over which the response coefficients change, the electromagnetic response of a medium can be described in terms of an effective medium theory, characterized by an effective translationally-invariant (non-local) optical conductivity tensor. The determination of this effective tensor in the presence of the CME is the central aim of this paper.

In general, the space-dependent response coefficients can be decomposed into the sums of their volume-averaged parts, denoted with overlines, and random parts with zero averages:
\begin{eqnarray}
  \sigma_{ab}(\rr,\w)&=&\overline\sigma_{ab}(\w)+\delta\sigma_{ab}(\rr,\w),\nonumber\\
  \gamma_{abc}(\rr,\w)&=&\overline{\gamma}_{ab}(\w)+\delta\gamma_{abc}(\rr,\w).
  \end{eqnarray}
Since the CME is a relatively weak effect, and spatial fluctuations of $\delta\gamma_{abc}$ will lead to even weaker effects, we set $\delta\gamma_{abc}\to 0$ in what follows.

It should be stressed that the variation of $\gamma_{abc}$ is inevitably present near sample boundaries; however, this variation does not play any role in the effective medium construction, and only affects the boundary conditions for electromagnetic waves scattering off a sample with CMEor natural optical activity~\cite{AgranovichYudson,HosurKerr2014}.

In what follows, we assume egrodic behavior for fluctuations of response coefficients, in the sense that volume averages for various quantities coincide with their ensemble averages over disorder realizations. Physically, this means that we neglect the mesoscopic fluctuations of effective medium parameters.

To construct the effective medium theory, we use the Maxwell equations to describe the sample-specific response of a disordered material to electromagnetic fields, and then average it over the disorder realizations.  The electromagnetic response of the medium is fully determined by its non-local optical conductivity tensor. To the lowest order in spatial gradients of the electric field, one has the following expression for the $a^{th}$ component of the current density in a non-uniform chiral metal:
\begin{equation}\label{eq:generalcurrent}
  j_a=\sigma_{ab}(\rr,\w)E_{b}+\frac{i}{\w}\gamma_{abc}(\w)\nabla_c E_b.
\end{equation}
This expression is the anisotropic version of Eq.~\eqref{eq:currentgeneral} in view of the Faraday's law for monochromatic fields, $\BB=\nbl\times \EE/i\w$. It is well known that, in a time-reversal system, the antisymmetric part of the optical conductivity tensor is fully determined by spatial gradients of $\gamma_{abc}$~\cite{AgranovichYudson}. Since we take $\gamma_{abc}(\w)$ to be equal to its spatially averaged (equally, disorder-averaged) value, $\sigma_{ab}(\rr,\w)$ is a symmetric local conductivity tensor.

To proceed, we make several simplifying assumptions, which are easily relaxed within the theory developed below, but increase the clarity of presentation. We go back to the assumption that the medium is isotropic, hence $\sigma_{ab}=\sigma\delta_{ab}$, $\gamma_{abc}=\gamma\e_{abc}$. Under these assumptions, the expression for the current density simplifies to
\begin{equation}
  \jj=(\sigma(\w)+\d\sigma(\w,\rr)) \EE+\gamma(\w) \BB.
\end{equation}

The fluctuations of the local conductivity tensor are assumed to be Gaussian, with a given correlator:
\begin{equation}
  \bra\d\sigma(\w,\rr) \d\sigma(\w,\rr')\ket=K_\w(\rr-\rr').
\end{equation}

The effective medium is then characterized by effective conductivity, $\sigma_{\textrm{eff}}$, and effective chiral magnetic conductivity, $\gamma_{\textrm{eff}}$, which relate the average current density to the average electric and magnetic fields:
\begin{equation}
  \overline \jj=\sigma_{\textrm{eff}}\overline \EE +\gamma_{\textrm{eff}}\,\,\overline \BB.
\end{equation}

Using the Maxwell equations, the electric field for a given realization of $\delta \sigma(\rr,\w)$ can be shown to satisfy
\begin{equation}
  \nbl(\nbl \EE)-\nbl^2\EE=\frac{\w^2}{c^2}\e(\w)\EE+\frac{\gamma(\w)\nbl\times \EE}{\e_0c^2}+\frac{i\w}{\e_0 c^2}\d\sigma(\w,\rr)\EE,
\end{equation}
with
\begin{equation}
  \e(\w)=\e_\infty+\frac{i\overline \sigma(\w)}{\e_0 \w}.
\end{equation}

The corresponding (retarded) Green's function obeys the following equation:
\begin{eqnarray}
  &&\left(\nabla_a\nabla_b-\nabla^2\d_{ab}-\frac{\w^2}{c^2}\e(\w)\d_{ab}+\frac{\gamma(\w)}{\e_0c^2}\e_{abd}\nabla_d\right.\nonumber\\
  &&\left.-\frac{i\w}{\e_0 c^2}\d\sigma(\w,\rr)\d_{ab}\right)D_{bc}(\rr,\rr';\w)=\d_{ac}\d(\rr-\rr').
\end{eqnarray}

For weak Gaussian disorder, the effective medium theory reduces~\cite{Kohler} to the standard self-consistent Born approximation for the disorder-averaged Green's function, $\overline D_{bc}(\rr-\rr',\w)$, which depends on the difference $\rr-\rr'$ due to the restored translational invariance.

Averaging over the ``disorder realizations'' is done according to the standard rules for systems with quenched disorder~\cite{abrikosov2012methods}. In particular, such averaging restores the translational invariance, and the medium is characterized by a self-energy in the expression for the average retarded Green's function of the electric fields. In the Fourier space, the equation for the disorder-averaged Green's function becomes
\begin{eqnarray}\label{eq:averageGF}
  &&\left(q^2\d_{ab}-q_aq_b-\frac{\w^2}{c^2}\e(\w)\d_{ab}+i\frac{\gamma(\w)}{\e_0c^2}\e_{abd}q_d\right.\nonumber\\
  &&-\Sigma_{ab}(\qq,\w)\Big)\overline D_{bc}(\qq,\w)=\d_{ac}.
\end{eqnarray}
The self-energy in real space is given by
\begin{equation}
  \Sigma_{ab}(\rr-\rr';\w)=-\frac{\w^2}{\e_0^2c^4} K_\w(\rr-\rr')\overline D_{ab}(\rr-\rr';\w),
\end{equation}
which can be rewritten in the Fourier space as
\begin{equation}\label{eq:SigmaFT}
  \Sigma_{ab}(\qq,\w)=-\frac{\w^2}{\e_0^2c^4}\int(d\qq') K_\w(\qq-\qq')\overline D_{ab}(\qq',\w).
\end{equation}
To capture the CME, one has to keep the linear in $\qq$ dependence of the self-energy. Due to the assumed isotropy of the medium, the latter can be decomposed as
\begin{equation}
  \Sigma_{ab}(\qq,\w)\approx \frac{\w^2}{c^2}\Sigma_0(\w)\d_{ab}-\frac{i}{\e_0c^2}\Sigma_1(\w)\e_{abc}q_c.
\end{equation}
From Eq.~\eqref{eq:averageGF} it is clear that $\Sigma_{0,1}$ play the role of corrections to the average values of $\e(\w)$ and $\gamma(\w)$, respectively. From~\eqref{eq:SigmaFT}, the expressions for $\Sigma_{0,1}$ read
\begin{eqnarray}
  &&\Sigma_0(\w)\d_{ab}=-\frac{1}{\e_0^2c^2}\int(d\qq) K_\w(\qq)\overline D_{ab}(\qq,\w),\nonumber\\
&&\Sigma_1(\w)\e_{abc}=\frac{i\w^2}{\e_0c^2}\int(d\qq) (\p_{q_c}K_\w(\qq))\overline D_{ab}(\qq,\w).
\end{eqnarray}
The fact that the tensor structures on left and right hand sides of these equations match is guaranteed by the isotropy of the medium.

Limiting ourselves to the linear order in $\gamma$, we finally obtain
\begin{eqnarray}\label{eq:Sigmas}
  &&\Sigma_0(\w)=
  \frac{1}{3\e_0^2c^2}\int(d\qq) K_\w(q)\left(\frac{2}{q_\w^2-q^2}+\frac1{q_{\w}^2}\right),\nonumber\\
  && \Sigma_1(\w)=\gamma(\w)\frac{\w^2}{3\e_0^2c^4}\int(d\qq)\frac{ q\p_{q}K_\w(q)}{(q^2-q_w^2)^2}.
\end{eqnarray}
where $q_\w^2=\frac{\omega^2}{c^2}(\e(\w)+\Sigma_0(\w))$. The effective medium parameters are given by
\begin{eqnarray}\label{eq:effectiveparameters}
  \sigma_{\textrm{eff}}(\w)&=&\overline{\sigma}(\w)-i\epsilon_0\w\Sigma_0(\w),\nonumber\\
  \gamma_{\textrm{eff}}(\w)&=&\gamma(\w)+\Sigma_1(\w).
\end{eqnarray}

Eqs.~\eqref{eq:Sigmas} and~\eqref{eq:effectiveparameters} are one of the central results of this paper. They allow to determine the effective medium parameters for any particular model characterized by a given correlator of optical conductivity fluctuations.

In the equation for $\Sigma_0$, the first term in round brackets describes the contribution from the fluctuations with two transverse polarizations, while the second one is the contribution from the longitudinal electric field fluctuations. The latter are dispersionless, since we did not include quadratic spatial dispersion ($O(q^2)$) terms in the dielectric tensor. In general, the contribution from the transverse modes is small in parameter $\w^2\ell^2/c^2\sim \ell^2/\lambda_0^2$, where $\ell$ is the scale of macroscopic inhomogeneity, and $\lambda_0$ is the wavelength of the light with frequency $\w$ in vacuum.

To illustrate the theory, we consider a simple textbook problem of calculating $\Sigma_0$ for an insulator with space-dependent dielectric constant: $\e(\rr)=\e+\d\e(\rr)$, for $\w\to 0$. Here we neglect self-consistency to obtain known perturbative results. In this case, $\delta\s(\rr,\w)=-i\w\delta\e(\rr)$, and from~\eqref{eq:Sigmas} we immediately obtain
\begin{equation}\label{eq:emulsion}
  \overline \e=\e-\frac{1}{3}\frac{\overline{\d\e(\rr)^2}}{\e},
\end{equation}
where $\overline{\d\e(\rr)^2}$ is the mean square fluctuation of $\delta\e(\rr)$. Eq.~\eqref{eq:emulsion} is a well known result for the effective dielectric constant of an emulsion~\cite{LL8}.

\textit{A model with short-ranged correlations --} To apply general expressions~\eqref{eq:Sigmas} to a non-trivial situation, we consider a  metal with low carrier density, being treated within the Drude model with spatially-dependent electron density. In practice, one may talk about a doped semiconductor, taking into account spatial fluctuations of the dopant density. We will show that ensuing spatial fluctuations of the optical conductivity result in plasmonic features in the frequency dependence of $\gamma_{\textrm{eff}}(\w)$.

Within the Drude model, the spatially-dependent  optical conductivity has the following form:
\begin{equation}
  \s(\rr,\w)=\frac{\e_0\w_p^2\tau(\rr)}{1-i\w\tau(\rr)}.
\end{equation}

We are interested in plasmonic features in $\gamma_{\textrm{eff}}(\w)$, hence we specialize to frequencies close to the average plasma edge, $\w_0$. For $\w_0\tau\gg1$, the conductivity can be approximated according to
\begin{equation}
  \sigma(\w,\rr)\approx \frac{i\e_0\w_p^2(\rr)}{\w}+\frac{\e_0\w_p^2(\rr)}{\w^2\tau(\rr)}.
\end{equation}
In what follows we will neglect the real part of conductivity, since dissipation ($\Im\Sigma_0\neq 0$) will be generated by wave decay into plasmons. However, the (positive) $\emph sign$ of the real part of the conductivity sets the sign of $\Im\Sigma_0$ (also positive), see below.

Wrting $\w_p^2(\rr)=\w_0^2+\d\w^2_p(\rr)$, with
\begin{equation}
  \bra \d\w^2_p(\rr)\d\w^2_p(\rr')\ket=w^4\exp\left(-\kappa(\rr-\rr')\right),
\end{equation}
we obtain
\begin{equation}
  K_\w(q)=-\frac{\e_0^2w^4}{\w^2}\frac{8\pi \kappa}{(q^2+\kappa^2)^2}.
\end{equation}
Applying Eqs.~\eqref{eq:Sigmas}, we obtain a self-consistent equation for $\Sigma_0(\w)$:
\begin{equation}
  \Sigma_0(\w)=-
  \frac{1}{3}\frac{w^4}{\w^4}\frac1{1-\frac{\w_0^2}{\w^2}+\Sigma_0(\w)}.
\end{equation}
For $|\w^2-\w_0^2|<2w^2/\sqrt{3}$ one has
\begin{eqnarray}
  \Re\Sigma_0(\w)&=&\frac{\w_0^2-\w^2}{2\w^2},\nonumber\\
  \Im \Sigma_0&=&\sqrt{\frac13 \frac{w^4}{\w^4}-\frac14\left(\frac{\w_0^2}{\w^2}-1\right)^2},
\end{eqnarray}
and for $|\w^2-\w_0^2|>2w^2/\sqrt{3}$:
\begin{eqnarray}
  \Re\Sigma_0(\w)&=&\frac{\w_0^2-\w^2}{2\w^2}+\frac{\sgn(\w^2-\w_0^2)}{2}\sqrt{\left(\frac{\w_0^2}{\w^2}-1\right)^2-\frac43 \frac{w^4}{\w^4}},\nonumber\\
  \Im \Sigma_0&=&0.
\end{eqnarray}
Here $\sgn(x)$ is the sign function.

Calcualting $\Sigma_1$, we get the following expression for $\gamma_{\textrm{eff}}(\w)$:
\begin{equation}\label{eq:sigma1}
  \gamma_{\textrm{eff}}(\w)=\gamma(\w)\left[1+\left(\frac{w}{c\kappa}\right)^4\frac{1}{(1-iq_\w/\kappa)^4}\right].
\end{equation}
As before, $q_\w^2=\frac{\omega^2}{c^2}(\e(\w)+\Sigma_0(\w))$.

\begin{figure}
  \centering
  \includegraphics[width=3.5in]{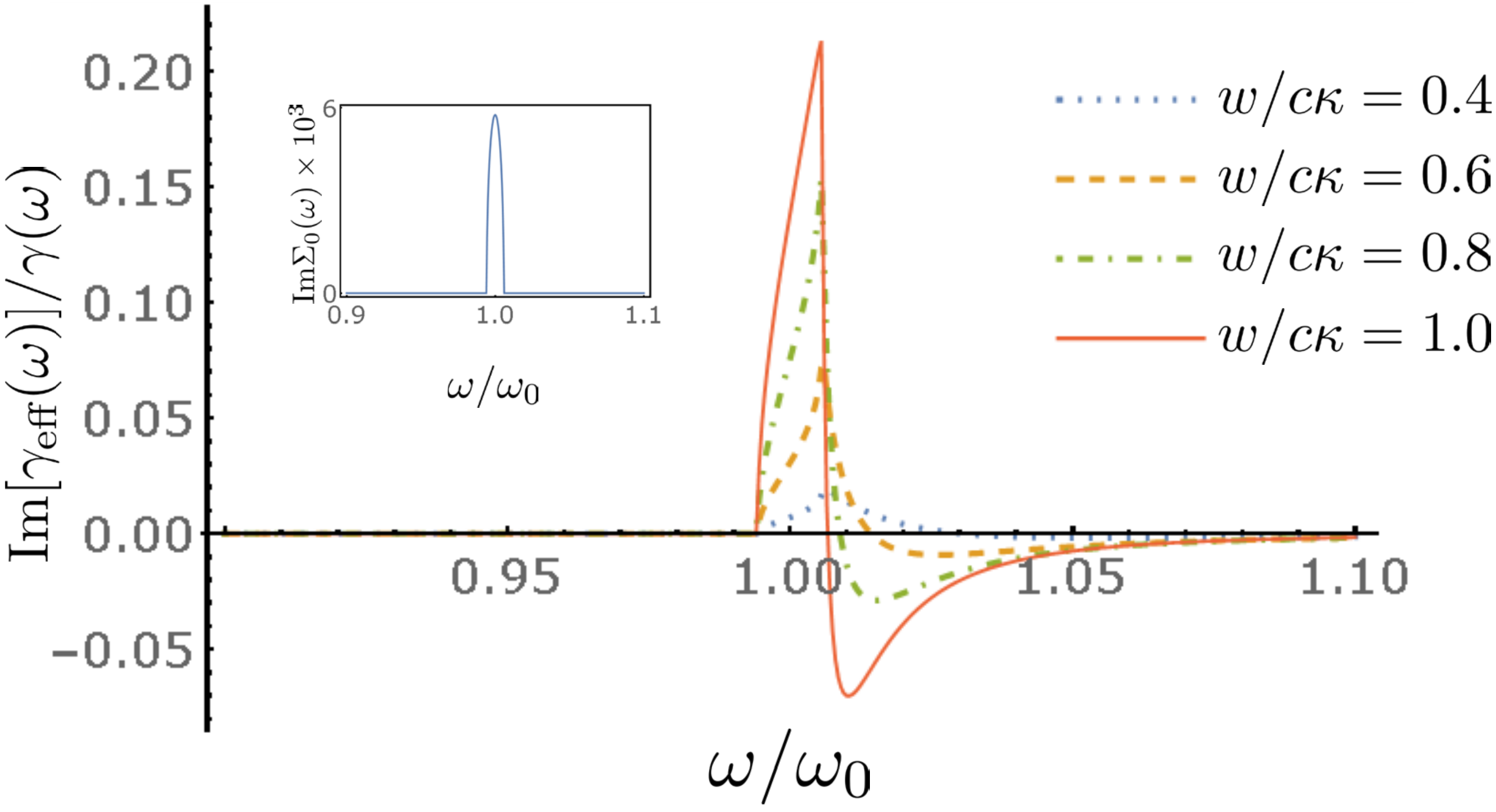}
  \caption{(Color online) Relative change in the imaginary part of the effective chiral conductivity for $w/\w_0=0.1$ and various values of $w/c\kappa$. Inset: Imaginary part of the disorder-induced correction to the dielectric function.}\label{fig:Im}
\end{figure}

\begin{figure}
  \centering
  \includegraphics[width=3.5in]{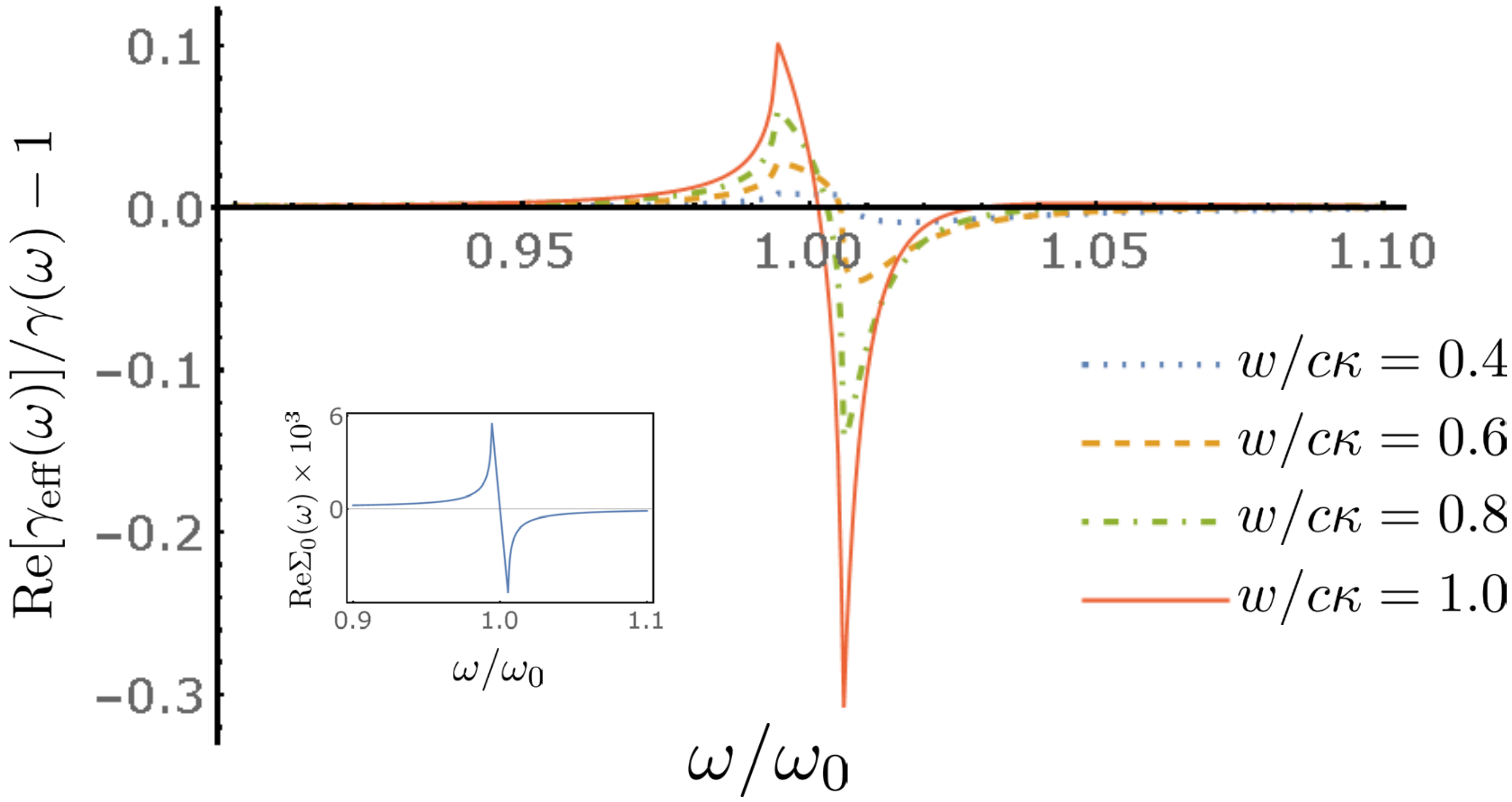}
  \caption{(Color online) Relative change in the rea; part of the effective chiral conductivity for $w/\w_0=0.1$ and various values of $w/c\kappa$. Inset: Real part of the disorder-induced correction to the dielectric function.}\label{fig:Re}
\end{figure}

The results of this calculation are plotted in Figs.~\ref{fig:Im} and~\ref{fig:Re}. It is observed that due to the disorder-induced scattering into the dispersionless plasmons the local part of the effective dielectric tensor of the medium acquires an imaginary part sharply peaked around the plasma frequency. In turn, this translates into sharp features in the circular dichroism and polarization rotation signals, which are determined by $\Im\gamma_{\textrm{eff}}(\w)$ and $\Re\gamma_{\textrm{eff}}(\w)$, respectively.

The results depend strongly on the values of two dimensionless parameters: $w/\w_0$, and $w/c\kappa$. It is hard to theoretically estimate them for a given material. Instead, they can be determined from the widths and maximum height of experimental peaks, analogous to those shown in Figs.~\ref{fig:Im} and~\ref{fig:Re}. For the particular model considered, the peak width scales as $w^2/\w^2_0$, while peak values of circular dichroism and polarization rotation signals scale roughly as $(w/c\kappa)^2$ and $(w/c\kappa)^4$ , respectively.

In summary, we have developed the theory of disorder-induced corrections to the chiral magnetic effect and natural optical activity in samples with macroscopic inhomogeneities. The theory is applicable to situations in which the electromagnetic fields vary smoothly on the inhomogeneity scale. In particular, the theory pertains to the case of Weyl metals with low electron density, in THz frequency range. In general, the disorder-induced correction are not large in absolute magnitude, but are the primary source of sharp frequency dependence of the chiral conductivity around the plasma edge of the metal. This observation is pertinent to any helical metals with natural optical activity, not just Weyl ones.

\acknowledgments{We would like to thank M. E. Raikh for useful discussions. This work was supported by NSF Grant No. DMR-1409089.}

\end{document}